\documentclass[twocolumn,showpacs,preprintnumbers,amsmath,amssymb,superscriptaddress,prl]{revtex4}

\usepackage{amsmath}

\usepackage{amssymb}

\usepackage{graphicx}

\usepackage{dcolumn}

\usepackage{bm}

\usepackage{color}

\def\be{\begin{eqnarray}}

\def\ee{\end{eqnarray}}

\def\ba{\begin{array}}

\def\ea{\end{array}}

\def\nn{\nonumber}

\newcommand{\un}[1]{\mathrm{#1}}

\newcommand{\rh}{R_{\rm H}}

\newcommand{\rl}{R_{\rm L}}

\begin{document}


\title{Interaction mediated asymmetries of the quantized--Hall--effect}

\author{A. Siddiki} %

\affiliation{Center for NanoScience and Fakult\"at f\"ur Physik,
Ludwig--Maximilians--Universit\"at, Geschwister--Scholl--Platz 1,
D--80539 M\"unchen, Germany} \affiliation{Physics Department,
Faculty of Arts and Sciences, 48170-Kotekli, Mugla, Turkey}



\author{J. Horas} %

\affiliation{Center for NanoScience and Fakult\"at f\"ur Physik,
Ludwig--Maximilians--Universit\"at, Geschwister--Scholl--Platz 1,
D--80539 M\"unchen, Germany}

\author{J. Moser} %

\affiliation{Center for NanoScience and Fakult\"at f\"ur Physik,
Ludwig--Maximilians--Universit\"at, Geschwister--Scholl--Platz 1,
D--80539 M\"unchen, Germany}

\author{W. Wegscheider} %
\affiliation{Institut f\"ur Experimentelle und Angewandte Physik,
Universit\"at Regensburg, D--93040 Regensburg, Germany}
\author{S. Ludwig} %
\affiliation{Center for NanoScience and Fakult\"at f\"ur Physik,
Ludwig--Maximilians--Universit\"at, Geschwister--Scholl--Platz 1,
D--80539 M\"unchen, Germany}

\begin{abstract}
Experimental and theoretical investigations on the integer
quantized--Hall--effect in gate defined narrow Hall--bars are
presented. At low electron mobility the classical (high
temperature) Hall--resistance line $\rh(B)$ cuts through the
center of all Hall--plateaus. In contrast, for our high mobility
samples the intersection point, at even filling factors
$\nu=2,4,\dots$, is clearly shifted towards larger magnetic fields
$B$. This asymmetry is in good agreement with predictions of the
screening theory, i.\,e.\ taking Coulomb--interaction into
account. The observed effect is directly related to the formation
of incompressible strips in the Hall--bar. The spin--split plateau
at $\nu=1$ is found to be almost symmetric regardless of the
mobility. We explain this within the so-called effective
$g$--model.
\end{abstract}




\maketitle

The integer quantized--Hall--effect (IQHE) can be observed when a
two dimensional electron system (2DES) at low temperature is
subjected to a strong magnetic field $B$ normal to the plane of
the 2DES. The relevance of the IQHE stems from its universal
features. Most prominent are the precise values $\rh= h/N e^2$
(with Planck's constant $h$, the elementary charge $e$, and a
natural number $N=1,2,\dots$) the quantized Hall--resistance takes
on the Hall--plateaus while at the same time the longitudinal
resistance $\rl$ vanishes \cite{vKlitzing80:494}. These
main--characteristics of the IQHE are well established in
experiments as well as within single--particle theories
\cite{vKlitzing80:494,Laughlin81,Buettiker86:1761}. However, these
conventional theories do not provide a full understanding of all
features observed in magneto--resistance experiments. A
comprehensive model needs to take into acount the
Coulomb--interaction between charge carriers
\cite{Chklovskii92:4026}, which is a subject of ongoing
investigations \cite{Gerhardts08:378,Sefa08:prb}.

In the classical (high temperature) limit the Hall--resistance
$\rh(B)$ resembles a straight line described by $\rh(B)=h/\nu_0(B)
e^2$ with $\nu_0(B)$ being the filling factor averaged across the
Hall--bar width. The local filling factor is defined as
$\nu(B,x)=n_{\rm s}(x,B)/n_{\phi}(x,B)$, where $n_{\rm s}$ and
$n_{\phi}\propto B$ are the local number densities of electrons
and magnetic flux quanta in the 2DES. In most experiments
reported, the Hall--plateaus of $\rh(B)$ extend symmetrically in
respect to integer values of $\nu_0\equiv N=1,2,\dots$. In other
words the classical Hall--line $\rh(B)$ cuts through the center of
each plateau \cite{Matthews05:497}. Exceptions from such symmetric
plateaus have been observed on etched narrow Hall--bars in the
limit of low mobility \cite{Zheng85:size,Haug87:SdH}. The
experimental results reported in Ref. \cite{Haug87:SdH} have been
described within single particle theories
\cite{Laughlin81,Buettiker86:1761} making additional assumptions
about the disorder potential, namely by comparison of the electron
diffusion length and the sample width \cite{Haug87:SdH}. In
earlier experiments asymmetric plateaus were attributed to
interactions~\cite{Zheng85:size}.

We present investigations on the IQHE as a function of mobility
and temperature employing narrow Hall--bars. Our devices are
electrostatically defined by top gates, allowing for very smooth
edges of the Hall--bar. The Hall--plateaus at even filling factors
develop a pronounced asymmetry while temperature is decreased.
This asymmetry is observed only at high mobilities, where the
electron mean--free path ($l_{\rm mfp}$) exceeds the Hall--bar
width. Hence, we can exclude disorder as the origin in contrast to
Ref.\ \cite{Haug87:SdH}. Considering the Coulomb--interaction
between electrons our results are qualitatively explained using
self--consistent (SC) calculations
\cite{siddiki2004,Siddiki:ijmp}. The model predicts an
interaction--induced asymmetric density of states (DOS) for charge
carriers within the Landau--levels \cite{Gerhardts87:asymmetry}.

\begin{figure}
{\centering
\includegraphics[width=1\linewidth]{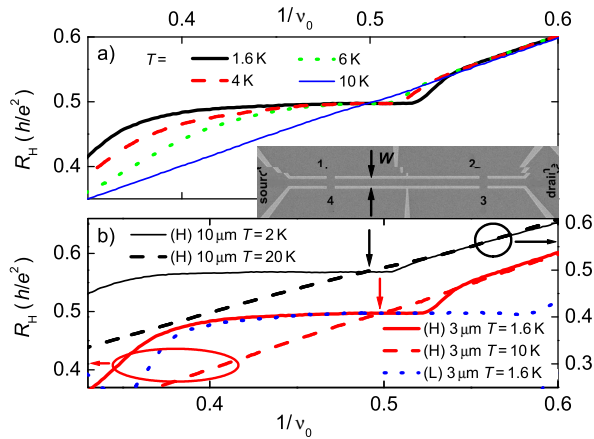}
\caption{ \label{fig:1} (color online) (a) Measured
Hall--resistance of the higher mobility wafer at a Hall--bar width
$W=3\,\mu$m for several temperatures as a function of the averaged
reciprocal filling factor at the $\nu_0=2$--plateaus. Inset:
Scanning electron micrograph of the gate layout. Metal gates are
light gray. Ohmic contacts source and drain carry the current
while 1--4 are voltage probes. (b) Hall--resistances in the limit
of high and low temperatures of the higher mobility wafer (H) at
$W=10\,\mu$m and $W=3\,\mu$m and of the lower mobility wafer (L)
at $W=3\,\mu$m.}}
\end{figure}

The experiments presented here are performed on two similar
GaAs/AlGaAs--heterostructures both containing a 2DES $110\,$nm
below the surface. The low temperature charge carrier densities
and mobilities of the two wafers are $n_{{\rm s}1}\simeq 2.8\times
10^{15}\,\un{m}^{-2}$, $n_{{\rm s}2}\simeq 1.8\times
10^{15}\,\un{m}^{-2}$, $\mu_1\simeq 140\,\un{m}^2/\un{Vs}$
($l_{\rm mfp}\approx 12\mu$m), and $\mu_2\simeq
300\,\un{m}^2/\un{Vs}$ ($l_{\rm mfp}\approx 21\mu$m). A typical
gate layout processed by electron beam lithography is displayed in
the inset to Fig.\ \ref{fig:1}a.
All gates of a sample are biased with the same negative voltage to
locally deplete the 2DES beneath the gates and thus define the
Hall--bar.  Measurements of the Hall--resistance $\rh$ are carried
out using the contacts 1--3 (or 2--4) as voltage probes. Likewise
contacts 1--2 (or 3--4) serve to measure the longitudinal
resistance $\rl$.

Fig.\ \ref{fig:1}a displays $\rh$ as a function of
$1/\nu_0\,(\propto B)$ in the $\nu_0\simeq 2$ range for
temperatures between $1.6\,{\mathrm K}\le T\le 10\,{\mathrm K}$
measured on the higher mobility wafer at a Hall--bar width of
$W=3\,\mu$m. At $T=10$\,K we find the classically expected
straight line and, as the temperature is decreased, the
Hall--plateau at $\nu_0=2$ develops. Noticeably, the plateau grows
stronger on the low magnetic field side ($\nu_0>2$), ultimately
resulting in an asymmetric plateau at low temperatures. Fig.\
\ref{fig:1}b displays $\rh(1/\nu_0)$ in the region of the
$\nu_0\simeq 2$ plateau measured on both wafers for $W=10\,\mu$m
and $W=3\,\mu$m at temperatures $T\lesssim 2\,$K as well as
$T\gtrsim 10\,$K. For the lower mobility wafer the classical high
temperature line cuts roughly through the center of the plateau as
expected in single particle models
\cite{Laughlin81,Buettiker86:1761}. In contrast, for the higher
mobility we again find asymmetric plateaus. This behavior is
likewise for larger even filling--factors (not shown). The main
experimental observations can be summarized as follows: i) In the
limit of high mobility Hall--plateaus in narrow gate defined bars
are asymmetric in respect to the classical $\rh$--line. ii) As the
mobility is reduced the conventional symmetric plateaus are
recovered. This makes disorder unlikely as possible origin of the
observed asymmetry. Instead, we consider the Coulomb--interaction
between electrons. In the following a SC model is briefly
introduced \cite{siddiki2004}. We start from the single particle
Hamiltonian but then explicitly include Coulomb interaction.

Consider an electron with charge $e$, effective mass $m^*$, and
momentum {\bf p} moving in a time--independent potential $V_{\rm
ext}(\textbf{r})$, generated by the top--gates as well as ionized
donors and other defects. In a magnetic field $B$ oriented
perpendicular to the 2DES described by the vector potential
$\textbf{A}(\textbf{r})$ (in an appropriate gauge) the Hamilton
operator reads
\be H=\frac{(\textbf{p}-e\textbf{A}(\textbf{r}))^2}{2m^*}+V_{\rm
ext}(\textbf{r})+V_{\rm e-e}(\textbf{r}) +\sigma g^*\mu_{\rm
B}B\,.\label{eq:hamilton}\ee
The potential $V_{e-e}(\textbf{r})$ accounts for
Coulomb--interactions between electrons with spin $\sigma$
$=\pm1/2$, where $g^*$ is the Lande--g--factor and $\mu_{\rm B}$
Bohr's magneton. We assume i) translational invariance in the
$y$--direction along the Hall--bar \cite{Lier94:7757}, ii) that
all charge carriers reside on the $z=0$ plane
\cite{Chklovskii92:4026}, iii) that disorder induces a mobility
dependent short range broadening of the DOS, $D(E)$, with
scattering parameter $\Gamma$ \cite{siddiki2004}, and iv) that the
electrostatic potential varies weakly on the scale of the magnetic
length $l_{\rm B}=\sqrt{\hbar/eB}$. Assumptions i) and ii) allow
to reduce the position vector to the lateral coordinate across the
Hall--bar ($\textbf{r}=(x,y,z)\rightarrow (x,y_0,0)\rightarrow
x$). We replace the actual wave functions of the electrons with
delta functions and apply the Thomas--Fermi--approximation
\cite{Guven03:115327} neglecting the spin degree of freedom
($g^*=0$) resulting in the carrier density
\be n_{\rm s}(x)=\int dE D(E)\left[e^{\frac{E-\mu(x)}{k_{\mathrm
B}T}}+1\right]^{-1}.\label{eq:tfadens}\ee
To obtain local conductivities we perform a spatial averaging over
the Fermi wavelength ($\sim 33$ nm) simulating the finite extent
of the wave functions, thus, relaxing the strict locality of our
model. The electrochemical potential $\mu(x)=\mu^{*}_{\rm
eq}-V(x)$ is composed of the equilibrium chemical potential
$\mu^{*}_{\rm eq}$ and the total potential energy, containing both
the Coulomb interaction between electrons and the external
potential defining the Hall--bar
\be V(x)=V_{\rm ext}(x)+V_{\rm e-e}(x)= \nn
\\ \frac{2e^2}{\kappa}\int_{-d}^{d}\left[n_0-n_{\rm s}(\tilde
x)\right]K(x,\tilde x)d\tilde x, \label{eq:tfapot} \ee
expressed via a Kernel $K(x,\tilde x)$ \cite{siddiki2004} such
that $V(-d)=V(d)=0$ (at the Hall--bar boundaries). Here $2d<W$ is
the reduced sample width, taking into account the lateral
depletion beneath the top--gates, $\kappa$ is the average
dielectric constant, and $n_0$ the constant (and homogeneous)
effective donor number density. Eqs.\ \ref{eq:tfadens} and
\ref{eq:tfapot} complete our SC problem, which we solve
iteratively to obtain electrostatic quantities, such as the local
electric field $\textbf{E}(x)$.

Assuming a constant current $I=\int_{-d}^{d}j_y(x,y) dx$ along the
Hall--bar, that is in $y$ direction, the local current density
$\textbf{j}(x)$ results from Ohm's law
\be \bf{\nabla}\mu(x)/e\equiv
\textbf{E}(x)=\hat{\rho}(x)\textbf{j}(x), \label{eq:ohms}\ee
where the resistivity tensor $\hat{\rho}(x)$ is obtained from the
DOS \cite{Guven03:115327,siddiki2004} and taking into account
short--range potential fluctuations \cite{Siddiki:ijmp}. From
$\bf{\nabla}\cdot\textbf{j}(x)=0$ and $\bf{\nabla} \times
\textbf{E}(x)=\textbf{0}$ and utilizing the translational
invariance one obtains
\be \begin{array}{cc}
  j_x=0, & E_y(x)= E_y^0\equiv I/\int_{-d}^{d}\,\frac{{dx}}{{\rho_{\rm L}(x)}}, \\\\
  j_y(x)=E_y^0/\rho_{\rm L}(x), & E_x(x)=E_y^0\rho_{\rm H}(x)/\rho_{\rm L}(x), \\
\end{array}\label{lol} \ee
where $\rho_{\rm L}(x)$ and $\rho_{\rm H}(x)$ are the diagonal and
off-diagonal entries of the resistivity tensor, respectively, and
$E_y^0$ is a constant electric field oriented in the
$y$--direction. For a given current eq.~\ref{lol} leads to the
global resistances
\be \rh=\frac{V_H}{I}=\frac{E_{y}^{0}}{I}
\int_{-d}^{d}dx\frac{\rho_H(x)}{\rho_L(x)},\quad
R_\mathrm{L}=\frac{2dE_y^0}{I}\label{eq:RHRL}\,, \ee
where the electron temperature enters via eq.\ \ref{eq:tfadens}.

\begin{figure} {\centering
\includegraphics[width=1\linewidth]{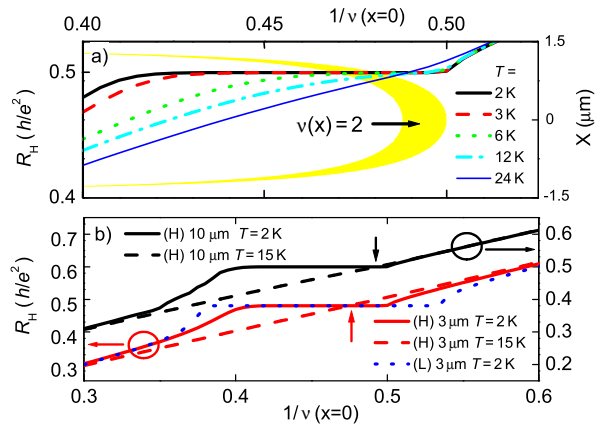}
\caption{ \label{fig:2} (color online) (a) Calculated
Hall-resistance for a high mobility (H) and $W=3\,\mu$m plotted
for several temperatures as a function of the reciprocal center
filling factor $1/\nu(x=0)$. The shaded region (yellow)
corresponds to the calculated spatial distribution of the IS with
$\nu(B,x)=2$ across the Hall--bar (rhs axis $0\le x\le W$) (b)
Hall--resistances as in Fig.\ \ref{fig:1}b but calculated assuming
that the biased gates result in an edge depletion of
$W/2-d=80\,$nm. The donor density is taken to be $4\times10^{15}$
m$^{-2}$, resulting in realistic Fermi energies of $E_\mathrm F =
11.9$\,meV ($13.4$\,meV) for $W=3\,\mu$m ($10\,\mu$m).}}
\end{figure}

Fig.\ \ref{fig:2}a
presents $\rh(B)$ of a Hall--bar of width $W=3\,\mu$m calculated
in the limit of high mobility (assuming a mean--free path large
compared to $2d\lesssim W$) as a function of a magnetic field
perpendicular to the 2DES for several temperatures $2\,{\rm K}\le
T \le 24\,{\rm K}$. Within an incompressible strip (IS) the
carrier density $n_{\rm s}(B,x)$ and, thus, the local filling
factor $\nu(B,x)$ are constant. In Fig.\ \ref{fig:2}a the IS with
$\nu(B,x)=2$ is highlighted depicted by a shaded region (yellow).
Here we display the bare IS neglecting broadening of the adjacent
compressible regions caused by temperature or the quantum
mechanical extension of the electron wave functions. At its high
magnetic field end (bulk region) the IS is extended over most of
the sample--width. As the $B$--field is reduced the IS splits into
two edge channels. Let us first consider the low temperature limit
of the local resistivity tensor. Away from the IS (white
background in Fig.\ \ref{fig:2}a) the compressible 2DES behaves
like a metal with finite diagonal elements $\rho_\mathrm L$ and
$\rho_\mathrm H$ taking a value close to its classical (high
temperature) limit. However, within an IS backscattering is
absent, hence $\rho_\mathrm L(\nu=N)=0$ and and
$\rho_H(\nu=N)=\frac{h}{Ne^2}$ takes its quantized value.
Accordingly, whenever somewhere across the Hall--bar an IS exists
$E_y^0=0$, and  eq.\ \ref{eq:RHRL} yields $R_\mathrm{L}=0$ and a
Hall--plateau with $\rh=\rho_H(\nu=N)=\frac{h}{Ne^2}$. The
calculated temperature dependence $R_{\rm H}(T)$ shown in Fig.\
\ref{fig:2}a is a consequence of the broadening of the
Fermi--distribution function with increasing temperature. Simply
speaking, a broader Fermi--distribution results in a wider
transition between compressible and incompressible regions,
melting an IS from its edges. Hence,  with increasing temperature
an IS and the according Hall--plateau disappear first where the
bare IS is narrow, hence on its low--magnetic field side (compare
Fig.\ \ref{fig:2}a). On the other hand the large (bulk) region of
an IS at its high magnetic field end withstands much higher
temperatures. As a direct consequence, the intersection point of
the classical (high temperature) $R_{\rm H}$--line with a
Hall--plateau is determined by the widest part of an IS.

Fig.\ \ref{fig:2}b displays calculated $R_{\rm H}$--curves as a
function of $1/\nu_0$ for the same two Hall--bar widths as the
actually measured devices have (compare Fig.\ \ref{fig:1}). For
the wider sample the two cases of a mean--free path much larger
(high mobility limit) or smaller (low mobility limit) than the
bar--width are presented. Long--range potential fluctuations
originating from charged impurities and resulting in a finite
mobility, are simulated by modulating the external potential
$V_\mathrm{ext}(x)\rightarrow
V_\mathrm{ext}(x)+V_\mathrm{mod}\cos(m_\mathrm p\pi x/d),
\label{eq:modulation}$ where $m_\mathrm p$ defines the mobility
\cite{Siddiki:ijmp}. For the low mobility limit we chose the
period $2d / m_\mathrm p \pi =1200\,\mathrm{nm}$ and a strong
modulation of $V_\mathrm{mod}\simeq E_\mathrm{F}/5$
\cite{Siddiki:ijmp}. The result are disorder broadened ISs
existing of bulk--regions extending more symmetrically in both
magnetic field directions around the integer filling factors
$\nu_0=N$. Consequently, for a low mobility also the
Hall--plateaus are almost symmetrically extended in respect to the
intersection point with the classical $R_{\rm H}$--line, being
independent on mobility.

The SC calculations presented in Fig.\ \ref{fig:2} show excellent
qualitative agreement with the measured data displayed in Fig.\
\ref{fig:1}. Our analysis indicates that the asymmetric
Hall--plateaus measured in the limit of high mobility and narrow
gate--defined Hall--bars, can be explained by the interaction
between charge carriers resulting in the formation of ISs. At high
mobility the Hall--resistance is quantized as long as there exists
an IS wide compared to the Fermi wave length. The long extension
of the measured Hall--plateaus to the low--field side of the
intersection with the classical Hall--line allows us to conclude,
that in narrow Hall--bars with high mobility and smooth (gate
defined) edges the edge potential profile rather than disorder
dominates the IQHE. When decreasing the mobility our measurements
and calculations show a transition to symmetric Hall--plateaus,
indicating that in this case disorder extends the large
bulk--region of the ISs in both field directions resulting in
symmetric plateaus. Our numerical calculations suggest that the
period $2d / m_\mathrm p \pi$ defines the long range length scale
of the disorder potential and, thus, compared to the sample width
is a measure of the mobility, suggesting a low mobility for $1 /
m_\mathrm p \pi \ll 1$.

\begin{figure}
{\centering
\includegraphics[width=0.9\linewidth]{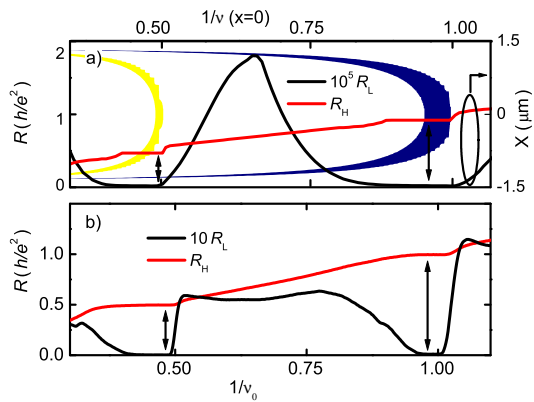}
\caption{ \label{fig:3} (color online) (a) Spatial distribution of
the ISs with $\nu(B,x)=1$ (dark, blue) and $\nu(B,x)=2$ (shaded
region, yellow) as a function of $1/\nu(x=0)$, calculated in the
effective $g$--factor model. Also shown are $\rh$-- and
$\rl$--curves calculated in the high mobility limit for $T=1.6\,$K
and $W=3\,\mu$m. (b) Measured $\rh(1/\nu_0)$ and $\rl(1/\nu_0)$
for the higher mobility wafer and $W=3\,\mu$m. The widest
extensions of the ISs are marked by arrows.}}
\end{figure}

It is known that exchange--correlation effects cause a spin--split
DOS usually expressed in a strongly enhanced effective g--factor
$g^*$ \cite{Khrapai:05}. This enhancement is expected to be even
aggravated within ISs. We include the spin degree of freedom in
our model in a phenomenological manner described in Ref.\
\cite{afifPHYSEspin}. While the exact value of $g^*$ is not a
determining parameter for our calculations, only a large enough
gap of the DOS $\Delta E_\mathrm Z\gg k_\mathrm B T$ results in
the formation of spin--split ISs. Fig.\ \ref{fig:3}a
presents the calculated spacial distributions of the bare ISs with
$\nu=1$ (dark, blue) and $\nu=2$ (light, yellow) together with
$\rh$ and $R_\mathrm L$ as a function of $1/\nu_0$. The
corresponding measured $\rh(1/\nu_0)$ and $\rl(1/\nu_0)$ curves
(for the higher mobility) are displayed in Fig.\ \ref{fig:3}b. To
calculate $\rl$ in the high mobility limit we used a
phenomenological model proposed by Gerhardts and Gross
\cite{Gross98:60}. An IS vanishes whenever the adjacent
compressible regions overlap, that is either when the quantum
mechanical wavelength of the electrons exceed its widths or when
the thermal energy $\sim k_{\mathrm B}T$ exceeds the local
potential drop. The latter is given by $g^*\mu_{\mathrm B} B$ for
$\nu=1$ or $\hbar\omega_\mathrm c -g^*\mu_{\mathrm B}B$ for even
filling factors. Where an IS exists $R_\mathrm{L}=0$ (and
$R_\mathrm{H}=const$). In our specific case the two ISs do not
coexist at any magnetic field value. The IS at $\nu=1$ is more
strongly developed and its bulk region extended over a larger
$B$--field interval compared to the IS at $\nu=2$. Arrows in Fig.\
\ref{fig:3} indicate the $1/\nu$--values where the ISs are widest,
i.\,e.\ where the classical Hall--line intersects with the
Hall--plateau. Clearly, the stronger developed IS at $\nu=1$
results in a more symmetric Hall--plateau compared to the even
filling factor $\nu=2$. Showing excellent agreement, the same
qualitative behavior is observed in the measured data in Fig.\
\ref{fig:3}b.

In conclusion, we have investigated the IQHE on gate defined
narrow Hall--bars at various mobilities and temperatures. At high
mobilities and low temperatures we observe asymmetric
Hall--plateaus in respect to the intersection point with the
classical Hall--resistance line. Our experimental findings are in
excellent agreement with predictions of the screening theory of
the IQHE. In contrast to the asymmetric plateaus at even filling
factors the measured spin-split plateau at $\nu=1$ is almost
symmetric. This is approved by model calculations within the
effective $g$--factor model.

We thank K.\,von Klitzing, R.\,R.\ Gerhardts, and K.\ Ikushima for
stimulating discussions. Financial support by the German Science
Foundation via the Germany Israel program DIP and the German
Excellence Initiative via the "Nanosystems Initiative Munich
(NIM)" is gratefully acknowledged.

\end{document}